
\font\gross=cmbx10  scaled\magstep2
\font\mittel=cmbx10 scaled\magstep1

\def\gsim{\mathrel{\raise.3ex\hbox{$>$\kern- .75em
                      \lower1ex\hbox{$\sim$}}}}
\def\lsim{\mathrel{\raise.3ex\hbox{$<$\kern-.75em
                      \lower1ex\hbox{$\sim$}}}}

\def\square{\kern 1pt
\vbox{\hrule height 0.6pt\hbox{\vrule width 0.6pt \hskip 3pt
\vbox{\vskip 6pt}\hskip 3pt\vrule width 0.6pt} \hrule height 0.6pt}\kern 1pt }

\def\sla{\raise.15ex\hbox{$/$}\kern-.72em}

\def\singlespace{\baselineskip=\normalbaselineskip}

\def\doublespace{\baselineskip=\normalbaselineskip \multiply\baselineskip by 2}
\parskip=\medskipamount
\overfullrule=0pt
\raggedbottom
\def\normalparindent{24pt}
\newif\ifdraft \draftfalse

\nopagenumbers
\footline={\ifnum\pageno=1 {\ifdraft
{\hfil\rm Draft \number\day -\number\month -\number\year}
\else{\hfil}\fi}
\else{\hfil\rm\folio\hfil}\fi}
\def\endpage{\vfill\eject}
\def\beginlinemode{\endmode\begingroup\parskip=0pt
\obeylines\def\\{\par}\def\endmode{\par\endgroup}}
\def\beginparmode{\endmode\begingroup \def\endmode{\par\endgroup}}
\let\endmode=\par
\def\raggedcenter{
                  \leftskip=2em plus 6em \rightskip=\leftskip
                  \parindent=0pt \parfillskip=0pt \spaceskip=.3333em
                  \xspaceskip=.5em\pretolerance=9999 \tolerance=9999
                  \hyphenpenalty=9999 \exhyphenpenalty=9999 }
\def\\{\cr}
\let\rawfootnote=\footnote\def\footnote#1#2{{\parindent=0pt\parskip=0pt
        \rawfootnote{#1}{#2\hfill\vrule height 0pt depth 6pt width 0pt}}}
\def\title{\null\vskip 3pt plus 0.2fill\beginlinemode\raggedcenter\gross}
\def\author{\vskip 3pt plus 0.2fill \beginlinemode\raggedcenter}

\def\abstract{\vskip 3pt plus 0.3fill \beginparmode{\noindent
{\mittel Abstract}:~}  }
\def\endtitlepage{\endpage\body}
\def\body{\beginparmode\parindent=\normalparindent}
\def\head#1{\par\goodbreak{\immediate\write16{#1}
      \vskip 0.4cm{\noindent\gross #1}\par}\nobreak\nobreak\nobreak\nobreak}

\def\finalcite{\citeall\ref\citeall\Ref}
\newif\ifannpstyle
\newif\ifprdstyle
\newif\ifplbstyle
\newif\ifwsstyle

\gdef\refto#1{\ifprdstyle  $^{\[#1] }$ \else
              \ifwsstyle$^{\[#1]}$  \else
              \ifannpstyle $~[\[#1] ]$ \else
              \ifplbstyle  $~[\[#1] ]$ \else
                                         $^{[\[#1] ]}$\fi\fi\fi\fi}
\gdef\refis#1{\ifprdstyle \item{~$^{#1}$}\else
              \ifwsstyle \item{#1.} \else
              \ifplbstyle\item{~[#1]} \else
              \ifannpstyle \item{#1.} \else
                              \item{#1.\ }\fi\fi\fi\fi }
\gdef\journal#1,#2,#3,#4.{
           \ifprdstyle {#1~}{\bf #2}, #3 (#4).\else
           \ifwsstyle {\it #1~}{\bf #2~} (#4) #3.\else
           \ifplbstyle {#1~}{#2~} (#4) #3.\else
           \ifannpstyle {\sl #1~}{\bf #2~} (#4), #3.\else
                       {\sl #1~}{\bf #2}, #3 (#4)\fi\fi\fi\fi}
\def\ref#1{Ref.~#1}
\def\Ref#1{Ref.~#1}
\def\cite#1{{#1}}\def\[#1]{\cite{#1}}

\def\(#1){(\call{#1})}
\def\call#1{{#1}}\def\taghead#1{{#1}}
\def\references{\head{References}\beginparmode\frenchspacing\parskip=0pt}
\def\endreferences{\body}
\def\endit{\endmode\vfill\supereject}\let\endpaper=\endit


\catcode`@=11
\newcount\r@fcount \r@fcount=0\newcount\r@fcurr
\immediate\newwrite\reffile\newif\ifr@ffile\r@ffilefalse
\def\w@rnwrite#1{\ifr@ffile\immediate\write\reffile{#1}\fi\message{#1}}
\def\writer@f#1>>{}
\def\referencefile{\r@ffiletrue\immediate\openout\reffile=\jobname.ref%
  \def\writer@f##1>>{\ifr@ffile\immediate\write\reffile%
    {\noexpand\refis{##1} = \csname r@fnum##1\endcsname = %
     \expandafter\expandafter\expandafter\strip@t\expandafter%
     \meaning\csname r@ftext\csname r@fnum##1\endcsname\endcsname}\fi}%
  \def\strip@t##1>>{}}

\def\citeall#1{\xdef#1##1{#1{\noexpand\cite{##1}}}}
\def\cite#1{\each@rg\citer@nge{#1}}
\def\each@rg#1#2{{\let\thecsname=#1\expandafter\first@rg#2,\end,}}
\def\first@rg#1,{\thecsname{#1}\apply@rg}
\def\apply@rg#1,{\ifx\end#1\let\next=\relax%
\else,\thecsname{#1}\let\next=\apply@rg\fi\next}%
\def\citer@nge#1{\citedor@nge#1-\end-}
\def\citer@ngeat#1\end-{#1}
\def\citedor@nge#1-#2-{\ifx\end#2\r@featspace#1
  \else\citel@@p{#1}{#2}\citer@ngeat\fi}
\def\citel@@p#1#2{\ifnum#1>#2{\errmessage{Reference range #1-#2\space is bad.}
    \errhelp{If you cite a series of references by the notation M-N, then M and
    N must be integers, and N must be greater than or equal to M.}}\else%
{\count0=#1\count1=#2\advance\count1 by1\relax\expandafter\r@fcite\the\count0,%
  \loop\advance\count0 by1\relax
    \ifnum\count0<\count1,\expandafter\r@fcite\the\count0,%
  \repeat}\fi}
\def\r@featspace#1#2 {\r@fcite#1#2,}    \def\r@fcite#1,{\ifuncit@d{#1}
    \expandafter\gdef\csname r@ftext\number\r@fcount\endcsname%
    {\message{Reference #1 to be supplied.}\writer@f#1>>#1 to be supplied.\par
     }\fi\csname r@fnum#1\endcsname}
\def\ifuncit@d#1{\expandafter\ifx\csname r@fnum#1\endcsname\relax%
\global\advance\r@fcount by1%
\expandafter\xdef\csname r@fnum#1\endcsname{\number\r@fcount}}
\let\r@fis=\refis   \def\refis#1#2#3\par{\ifuncit@d{#1}%
    \w@rnwrite{Reference #1=\number\r@fcount\space is not cited up to now.}\fi%
  \expandafter\gdef\csname r@ftext\csname r@fnum#1\endcsname\endcsname%
  {\writer@f#1>>#2#3\par}}
\def\r@ferr{\endreferences\errmessage{I was expecting to see
\noexpand\endreferences before now;  I have inserted it here.}}
\let\r@ferences=\references
\def\references{\r@ferences\def\endmode{\r@ferr\par\endgroup}}
\let\endr@ferences=\endreferences
\def\endreferences{\r@fcurr=0{\loop\ifnum\r@fcurr<\r@fcount
    \advance\r@fcurr by 1\relax\expandafter\r@fis\expandafter{\number\r@fcurr}%
    \csname r@ftext\number\r@fcurr\endcsname%
  \repeat}\gdef\r@ferr{}\endr@ferences}
\let\r@fend=\endpaper\gdef\endpaper{\ifr@ffile
\immediate\write16{Cross References written on []\jobname.REF.}\fi\r@fend}
\catcode`@=12
\finalcite
\catcode`@=11
\newcount\tagnumber\tagnumber=0
\immediate\newwrite\eqnfile\newif\if@qnfile\@qnfilefalse
\def\write@qn#1{}\def\writenew@qn#1{}
\def\w@rnwrite#1{\write@qn{#1}\message{#1}}
\def\@rrwrite#1{\write@qn{#1}\errmessage{#1}}
\def\taghead#1{\gdef\t@ghead{#1}\global\tagnumber=0}
\def\t@ghead{}\expandafter\def\csname @qnnum-3\endcsname
  {{\t@ghead\advance\tagnumber by -3\relax\number\tagnumber}}
\expandafter\def\csname @qnnum-2\endcsname
  {{\t@ghead\advance\tagnumber by -2\relax\number\tagnumber}}
\expandafter\def\csname @qnnum-1\endcsname
  {{\t@ghead\advance\tagnumber by -1\relax\number\tagnumber}}
\expandafter\def\csname @qnnum0\endcsname
  {\t@ghead\number\tagnumber}
\expandafter\def\csname @qnnum+1\endcsname
  {{\t@ghead\advance\tagnumber by 1\relax\number\tagnumber}}
\expandafter\def\csname @qnnum+2\endcsname
  {{\t@ghead\advance\tagnumber by 2\relax\number\tagnumber}}
\expandafter\def\csname @qnnum+3\endcsname
  {{\t@ghead\advance\tagnumber by 3\relax\number\tagnumber}}
\def\equationfile{\@qnfiletrue\immediate\openout\eqnfile=\jobname.eqn%
  \def\write@qn##1{\if@qnfile\immediate\write\eqnfile{##1}\fi}
  \def\writenew@qn##1{\if@qnfile\immediate\write\eqnfile
    {\noexpand\tag{##1} = (\t@ghead\number\tagnumber)}\fi}}
\def\callall#1{\xdef#1##1{#1{\noexpand\call{##1}}}}
\def\call#1{\each@rg\callr@nge{#1}}
\def\each@rg#1#2{{\let\thecsname=#1\expandafter\first@rg#2,\end,}}
\def\first@rg#1,{\thecsname{#1}\apply@rg}
\def\apply@rg#1,{\ifx\end#1\let\next=\relax%
\else,\thecsname{#1}\let\next=\apply@rg\fi\next}
\def\callr@nge#1{\calldor@nge#1-\end-}\def\callr@ngeat#1\end-{#1}
\def\calldor@nge#1-#2-{\ifx\end#2\@qneatspace#1 %
  \else\calll@@p{#1}{#2}\callr@ngeat\fi}
\def\calll@@p#1#2{\ifnum#1>#2{\@rrwrite{Equation range #1-#2\space is bad.}
\errhelp{If you call a series of equations by the notation M-N, then M and
N must be integers, and N must be greater than or equal to M.}}\else%
{\count0=#1\count1=#2\advance\count1 by1\relax\expandafter\@qncall\the\count0,%
  \loop\advance\count0 by1\relax%
    \ifnum\count0<\count1,\expandafter\@qncall\the\count0,  \repeat}\fi}
\def\@qneatspace#1#2 {\@qncall#1#2,}
\def\@qncall#1,{\ifunc@lled{#1}{\def\next{#1}\ifx\next\empty\else
  \w@rnwrite{Equation number \noexpand\(>>#1<<) has not been defined yet.}
  >>#1<<\fi}\else\csname @qnnum#1\endcsname\fi}
\let\eqnono=\eqno\def\eqno(#1){\tag#1}\def\tag#1$${\eqnono(\displayt@g#1 )$$}
\def\aligntag#1\endaligntag  $${\gdef\tag##1\\{&(##1 )\cr}\eqalignno{#1\\}$$
  \gdef\tag##1$${\eqnono(\displayt@g##1 )$$}}
\def\eqalignno#1{\displ@y \tabskip\centering
  \halign to\displaywidth{\hfil$\displaystyle{##}$\tabskip\z@skip
    &$\displaystyle{{}##}$\hfil\tabskip\centering
    &\llap{$\displayt@gpar##$}\tabskip\z@skip\crcr
    #1\crcr}}
\def\displayt@gpar(#1){(\displayt@g#1 )}
\def\displayt@g#1 {\rm\ifunc@lled{#1}\global\advance\tagnumber by1
        {\def\next{#1}\ifx\next\empty\else\expandafter
        \xdef\csname @qnnum#1\endcsname{\t@ghead\number\tagnumber}\fi}%
  \writenew@qn{#1}\t@ghead\number\tagnumber\else
        {\edef\next{\t@ghead\number\tagnumber}%
        \expandafter\ifx\csname @qnnum#1\endcsname\next\else
        \w@rnwrite{Equation \noexpand\tag{#1} is a duplicate number.}\fi}%
  \csname @qnnum#1\endcsname\fi}
\def\eqnoa(#1){\global\advance\tagnumber by1\multitag{#1}{a}}
\def\eqnob(#1){\multitag{#1}{b}}
\def\eqnoc(#1){\multitag{#1}{c}}
\def\eqnod(#1){\multitag{#1}{d}}
\def\multitag#1#2$${\eqnono(\multidisplayt@g{#1}{#2} )$$}
\def\multidisplayt@g#1#2 {\rm\ifunc@lled{#1}
        {\def\next{#1}\ifx\next\empty\else\expandafter
        \xdef\csname @qnnum#1\endcsname{\t@ghead\number\tagnumber b}\fi}%
  \writenew@qn{#1}\t@ghead\number\tagnumber #2\else
        {\edef\next{\t@ghead\number\tagnumber #2}%
        \expandafter\ifx\csname @qnnum#1\endcsname\next\else
    \w@rnwrite{Equation \noexpand\multitag{#1}{#2} is a duplicate number.}\fi}%
  \csname @qnnum#1\endcsname\fi}
\def\ifunc@lled#1{\expandafter\ifx\csname @qnnum#1\endcsname\relax}
\let\@qnend=\end\gdef\end{\if@qnfile
\immediate\write16{Equation numbers written on []\jobname.EQN.}\fi\@qnend}
\catcode`\|=13

\catcode`\;=13

\def\table{
     \begingroup
       \catcode`\|=13
       \catcode`\;=13
       \def|{&\I&&} 
       \def;{&&&}   
       \def\II{\vrule width 2.0pt}
       \def\I{\vrule}
       \def~{\hfill}
       \def\\{\strut\cr}
       \def\>{\hbox{\vbox to 14pt{}\vtop to 9pt{}}\cr}
       \def\space##1{\vbox to ##1pt{}}
       \def\----{\noalign{\hrule}\cr}
       \def\===={\noalign{\hrule height 2.0pt}\cr}
       \def\thinline{&\multispan3\hrulefill}
       \def\thickline{&\multispan3{\leaders\hrule height 2.0pt\hfill}}
       \def\noline{&\multispan2&}
       \def\makeline##1{##1\cr}
       \def\[{\II&&} 
       \def\]{&\II \>} 
       \def\({&&} 
       \def\){&\\} 
     \tableaux}
\def\tableaux#1#2{\vbox{
      \tabskip=0pt
      \offinterlineskip
      \parskip=0pt
     \halign to #1{&\tabskip=0pt##& ##\tabskip=0.5em plus 4em minus0.5em
     &\hfil##\hfil\cr
      #2}}
    \endtable}

\def\endtable{\catcode`\|=12 \catcode`\;=12\endgroup}
\catcode`\|=12
\catcode`\;=12
\magnification=1200
\baselineskip=12pt
\title
STRING INSTABILITIES
\title
IN BLACK HOLE SPACETIMES
\bigskip\bigskip\bigskip\bigskip
\author

C. O. LOUST\' O\footnote{$^{a}$}{
Universit\"at Konstanz, Fakult\"at f\"ur Physik, Postfach 5560, D-7750,
Konstanz, Germany. e-mail:
phlousto@dknkurz1.}{$^,$}\footnote{$^{b}$}{
Permanent Address: IAFE, Cas. Corr. 67, Suc. 28, 1428 Buenos Aires, ARGENTINA.
e-mail: lousto@iafe.edu.ar.}
and
N. S\' ANCHEZ\footnote{$^c$}{
Observatoire de Paris, Section Meudon, Demirm, UA 336 Laboratoire associ\' e
au CNRS, Observatoire de Meudon et \' Ecole Normale Sup\' erieure, 92195 Meudon
Principal Cedex, FRANCE.}

\bigskip\bigskip\bigskip\bigskip
\bigskip\bigskip
\noindent
\bigskip\bigskip
\endtitlepage
\vfill\eject
\singlespace
\centerline{\bf ABSTRACT}

We study the emergence of string instabilities in $D$ - dimensional black
hole spacetimes (Schwarzschild and Reissner - Nordstr\o m), and De Sitter
space (in static coordinates to allow a better comparison with the black
hole case). We solve the first order string fluctuations around the center
of mass motion at spatial infinity, near the horizon and at the spacetime
singularity. We find that the time components are always well behaved in
the three regions and in the three backgrounds. The radial components are
{\it unstable}: imaginary frequencies develop in the oscillatory modes near
the horizon, and the evolution is like $(\tau-\tau_0)^{-P}$, $(P>0)$, near
the spacetime singularity, $r\to0$, where the world - sheet time
$(\tau-\tau_0)\to0$, and the proper string length grows infinitely. In the
Schwarzschild black hole, the angular components are always well - behaved,
while in the Reissner - Nordstr\o m case they develop instabilities inside
the horizon, near $r\to0$ where the repulsive effects of the charge dominate
over those of the mass. In general, whenever large enough repulsive effects
in the gravitational background are present, string instabilities develop.
In De Sitter space, all the spatial components exhibit instability. The
infalling of the string to the black hole singularity is like the motion of a
particle in a potential $\gamma(\tau-\tau_0)^{-2}$ where $\gamma$ depends on
the $D$ spacetime dimensions and string angular momentum, with $\gamma>0$
for Schwarzschild and $\gamma<0$ for Reissner - Nordstr\o m black holes.
For $(\tau-\tau_0)\to0$ the string ends trapped by the black hole singularity.

\singlespace
\vfill\eject

\head{I - Introduction}

The study of the string dynamics in curved space-times, reveals new insights
with respect to string propagation in flat space-time (see for example refs
[\cite{vs87,vs88,sv90,gsv,vs91,vms}]).

The equations of motion and constraints for strings in curved spacetimes
are highly non linear (and, in general, not exactly solvable). In ref
[\cite{vs87}], a method was proposed (the ``strong field expansion") to
study systematically ( and approximately), the string dynamics in the
strong curvature regime. In this method, one starts from an exact particular
solution of the string equations in a given metric and then, one constructs
a perturbative series around this solution. The space of solutions for the
string coordinates is represented as
$$X^A(\sigma,\tau)=q^A(\sigma,\tau)+\eta^A(\sigma,\tau)+\xi^A(\sigma,\tau)+
.....~~,\eqno(i1)$$
$A=0,...,D-1$. Here $q^A(\sigma,\tau)$ is an exact solution of the string
equations and $\eta^A(\sigma,\tau)$ obeys a linearized perturbation around
$q^A(\sigma,\tau)$. $\xi^A(\sigma,\tau)$ is a solution of second perturbative
order around $q^A(\sigma,\tau)$. Higher order perturbations can be considered
systematically. A physically appealing starting solution is the center of
mass motion of the string, $q^A(\tau)$, that is, the point particle (geodesic)
motion. The world sheet time variable appears here naturally identified with
the proper time of the center of mass trajectory. The space time geometry is
treated {\it exactly}, and the string fluctuations around $q^A$ are treated
as perturbations. Even at the level of the zeroth order solution, gravitational
effects including those of the singularities of the geometry are fully taken
into account. This expansion corresponds to low energy excitations of the
string as compared with the energy associated to the geometry. This corresponds
to an expansion in powers of $(\alpha')^{1/2}$. Since $\alpha'=(l_{Planck})^2$,
the expansion parameter turns out to be the dimensionless constant
$$g=l_{Planck}/R_c=1/(l_{Planck}M)~~,\eqno(i2)$$
where $R_c$ characterizes the spacetime curvature and $M$ is its associated
mass
(the black hole mass, or the mass of a closed universe in cosmological
backgrounds). The expansion is well suited to describe strings in strong
gravitational regimes (in most of the interesting situations one has clearly
$g\ll1$). The constraint equations are also expanded in perturbations. The
classical $({\rm mass})^2$ of the string is defined through the center of
mass motion (or Hamilton - Jacobi equation). The conformal generators (or
world - sheet two dimensional energy - momentum tensor) are bilinear in the
fields $\eta^A(\sigma,\tau)$. [If this method is applied to flat spacetime,
the zeroth order plus the first order fluctuations provide the exact solution
of the string equations].

This method was first applied to cosmological (De Sitter) spacetimes. One of
the results was that for large enough Hubble constant, the frequency of the
lower string modes, i.e. those with $|n|<\alpha' mH$, ($\alpha'$ being the
string tension and $m$ its mass), becomes imaginary. This was further analysed
\refto{sv90,gsv} as the onset of a physical instability, in which the
proper string size starts to grow (precisely like the expansion factor of the
universe). The string modes couple with the background geometry in such a way
that the string inflates with the universe itself. The same happens for
strings in singular gravitational plane waves\refto{vs91,vms} (see
also ref[\cite{prd92}]), and the
results of this paper here show that this is a generic feature of strings
near spacetime singularities.

For black hole spacetimes, such unstable features had not been yet explored.
The string dynamics in black hole spacetimes is much more complicated to
solve ( even asymptotically and approximately). In ref [\cite{vs88}], the
study of string dynamics in a Schwarzschild black hole was started and the
scattering problem was studied for large impact parameters. Stable oscillatory
behavior of the string was found for the transversal (angular) components;
scattering amplitudes, cross section and particle transmutation process were
described, and explicitly computed in an expansion in $(R_s/b)^{D-3}$, $R_s$
being the Schwarzschild radius and $b$ the impact parameter. The aim of this
paper (and of a subsequent one\refto{LS93}), is to find, and then to describe,
the {\it unstable sector} of strings in black hole backgrounds. By unstable
behavior, we mean here the following characteristic features: non oscillatory
behavior in time, or the emergence of imaginary frequencies for some modes,
accompanied of an infinite stretching of the proper string length. In addition,
the spatial coordinates (some of its components) can become unbounded.
Stable string behavior means the usual oscillatory propagation with real
frequencies, (and the usual mode - particle interpretation), the fact that
the proper string size does not blow up, and that the string modes remain
well - behaved.

We express the first order string fluctuations $\eta^\mu$, $(\mu=0,....,D-1)$
in $D$ - dimensional Reissner - Nordstr\o m - De Sitter spacetime, as a
Schr\"odinger type equation for the amplitudes $\Sigma^\mu=q^R\eta^\mu$,
$q^R$ being the radial center of mass coordinate. We find the asymptotic
behavior of the longitudinal and transverse string coordinates ($\Sigma^+,
\Sigma^-,\Sigma^i$) with $i=2,.....,D-1$, at the spatial infinity, near the
horizon and near the spacetime singularity. $+$ and $-$ stand for the
longitudinal (temporal and radial) components respectively, and $i$ for the
transverse (angular) ones. We analyse first a head - on collision
(angular momentum $L=0$), that is, a radial infall of the string towards
the black hole. Then, we analyse the full $L\not=0$ situation. We consider
Schwarzschild, Reissner - Nordstr\o m and De Sitter spacetimes (described
here in static coordinates which allow a better comparison among the three
cases). In all the situations (with and without angular momentum) and for the
three cases we find the following results:

The time component $\Sigma^+$ is {\it  {always stable}} in the three
regions (near infinity, the horizon and the singularity), and in
the three cases (black holes and De Sitter spacetimes).

The radial component $\Sigma^-$ is {\it  {always unstable}} in the three
regions and in the three backgrounds. In the Schwarzschild case, the
instability condition for the radial modes - which develop imaginary
frequencies near the horizon - can be expressed as
$$n<{\alpha' m \sqrt{D-3}\over R_s}\left[D-2-\left({D-3\over2}\right)
{m^2\over E^2}\right]^2~,\eqno(i3)$$
where $\alpha'$, $m$ and $E$ are the string tension, string mass and energy
respectively. The quantity within the square brackets is always positive,
thus the lower modes develop imaginary frequencies when the typical string size
$\alpha' m \sqrt{D-3}$ is larger than the horizon radius. Notice the
similarity with the instability condition in De Sitter space,
$n<\alpha' m/r_H$, $r_H$ being the horizon radius.

In the Schwarzschild black hole, the transverse modes $\Sigma^i$ are stable
(well behaved) everywhere including the spacetime singularity at $q^R=0$.
In the Reissner - Nordstr\o m (R-N) black hole, the transverse modes
$\Sigma^i$ are stable at infinity and outside the horizon. Imaginary
frequencies appear, however, inside a region from $r_-<q^R<r_+$ to $q^R\to0$,
where $r_\pm=M\pm \sqrt{M^2-Q^2}$, $M$ and $Q$ being the mass and charge of the
black hole respectively. For the extreme black hole $(Q=M)$, instabilities
do not appear. There is a critical value of the electric charge of a
Reissner - Nordstr\o m black hole, above which the string passing through
the horizon passes from unstable to stable regime. In the De Sitter
spacetime, the only stable mode is the temporal one $(\Sigma^+)$. All the
spatial components exhibit instability, in agreement with the previous results
in the cosmological context\refto{vs87,sv90,gsv}. A summary of this analysis
is given in table 1.

Imaginary frequencies in the transverse string coordinates $(\Sigma^i)$
appear in the case in which the local gravity, i.e. $\partial_r a/2$, is
negative (that is, repulsive effects). Here,
$$a(r)=1-(R_s/r)^{D-3}+(\tilde Q^2/r^2)^{D-3}+
{\Lambda\over 3}r^2~,\eqno(i5)$$
where $\tilde Q^{2(D-3)}={8\pi GQ^2\over(D-2)(D-3)}$, and $\Lambda$ is the
cosmological constant. That is why the transverse modes $(\Sigma^i)$ are
well behaved in the Schwarzschild case, and outside the Reissner - Nordstr\o m
event horizon.
But close to $q^R\to0$, $a'_{RN}<0$ (Reissner - Nordstr\o m has a repulsive
inner horizon),
and the gravitational effect of the charge overwhelms that of the mass; in
this case instabilities develop. In the Reissner - Nordstr\o m -
De Sitter spacetime, unstable string behavior appears far away from the
black hole where De Sitter solution dominates, and inside the black hole
where the Reissner - Nordstr\o m solution dominates.
For $M=0$ and $Q=0$, we recover the instability
criterion\refto{vs87,sv90} $\alpha'm\Lambda/6>1$ for large enough Hubble
constant (this is in agreement with the criterion given in ref[\cite{g92}].

We find that in the black hole spacetimes, the transversal first order
fluctuations $(\Sigma^i)$ near the space time singularity $q^R=0$, obey
a Schr\"odinger type equation (with $\tau$ playing the role of a spatial
coordinate), with a potential $\gamma(\tau-\tau_0)^{-2}$,
(where $\tau_0$ is the proper time of arrival to the singularity at $q^R=0$).
 The dependence on
$D$ and $L$ is concentrated in the coefficient $\gamma$. Thus, the approach
to the black hole singularity is like the motion of a particle in a
potential $\gamma(\tau-\tau_0)^{-\beta}$, with $\beta=2$. And, then, like
the case $\beta=2$ of strings in singular gravitational waves\refto{vs91,vms}
(in which case the spacetime is simpler and the exact full string equations
become linear). Here $\gamma>0$ for strings in the Schwarzschild spacetime,
for which we have regular solutions $\Sigma^i$ ; while $\gamma<0$ for
Reissner - Nordstr\o m,
that is, in the case we have a singular potential and an unbounded behavior
(negative powers in $(\tau-\tau_0)$) for  $\Sigma^i_{RN}$.
The fact that the angular
coordinates $\Sigma^i_{RN}$ become unbounded means that the string makes
infinite turns around the spacetime singularity and remains trapped by it.

For $(\tau-\tau_0)\to0$, the string is trapped by
the black hole singularity. In
Kruskal coordinates $\left(u_k(\sigma,\tau), v_k(\sigma,\tau)\right)$, for
the Schwarzschild black hole we find
$\lim_{(\tau-\tau_0)\to0}u_kv_k=\exp{2KC(\sigma)(\tau-\tau_0)^P}$,
 where $K=(D-3)/(2R_s)$
is the surface gravity, $P>0$ is a
determined coefficient that depends on the $D$ dimensions, and $C(\sigma)$
is determined by the initial state of the string. Thus $u_kv_k\to1$ for
$(\tau-\tau_0)\to0$. The proper spatial string length at fixed
$(\tau-\tau_0)\to0$ grows like $(\tau-\tau_0)^{-(D-1)P}$.

It must be noticed that in cosmological inflationary backgrounds, the
unstable behavior manifests itself as non - oscillatory
in $(\tau-\tau_0)$ (exponential
for $(\tau-\tau_0)\to\infty$, power - like
for $(\tau-\tau_0)\to0$); the string coordinates
$\eta^i$ are constant (i.e. functions of $\sigma$ only), while the proper
amplitudes $\Sigma^i$ grow like the expansion factor of the universe.
In the black hole cases, and
more generally, in the presence of spacetime singularities, all the
characteristic features of string instability appear, but in addition the
spatial coordinates $\eta^i$ (or some of its components) become unbounded.
That is, not only the amplitudes $\Sigma^i$ diverge, but also the string
coordinates $\eta^i$, what appears as a typical feature of strings near
the black hole singularities. A full description of the string behavior
near the black hole singularity will be reported elsewhere\refto{LS93}.
This paper is organized as follows: In Section II we formulate the problem
of string fluctuations in the Reissner - Nordstr\o m - De Sitter
spacetime and express the
equations of motion in a convenient Sch\"odinger - type equation. In Section
III we treat the head - on collision. In Section IV and V we describe the
full $\tau$ dependence and the non - colinear case, respectively, and discuss
the conclusions and consequences of our results.

\head{II - Formulation of the problem}

de Vega and S\'anchez\refto{vs87}
have obtained the equations of motion
of fundamental strings in curved
 backgrounds by expanding the fluctuations
of the string around a given particular
solution of the problem (for example, the center of mass
motion). For the case of a black hole background\refto{vs88}
with mass $M$, charge $Q$ and cosmological constant $\Lambda$,
$$ds^2=-a(r)(dX^0)^2+a^{-1}(r)dr^2+r^2d\Omega^2_{D-2}~,$$
$$a(r)=1-(R_s/r)^{D-3}+(\tilde Q^2/r^2)^{D-3}+
{\Lambda\over 3}r^2~,\eqno(1')$$
$$R_s^{D-3}={16\pi GM\over(D-2)\Omega_{D-1}}~~,
{}~~\Omega_D={2\pi^{D/2}\over
\Gamma(D/2)}~~,~~\tilde Q^{2(D-3)}={8\pi GQ^2\over(D-2)(D-3)}~,$$
the equations of motion of the first order fluctuations read
$$\left[{d^2\over d\tau^2}+n^2\right] \eta_n^A(\tau)+
2A^A_B(\tau)\dot \eta^B_n(\tau)
+B^A_B(\tau)\eta^B_n(\tau)=0~.\eqno(1)$$
Here we have expanded the first order string perturbations
in a Fourier transform
$$\eta^A(\sigma,\tau)=\sum_n e^{in\sigma}\eta_n^A(\tau)~.\eqno(2)$$
$\eta^A$ being the vector
$$\eta=\pmatrix{\eta^0\cr \eta^*\cr \eta^i\cr}~~,
{}~~i=2,3,.....,D-1\eqno(3)$$
and $A^A_B(\tau)$ and $B^A_B(\tau)$ are the components of the following
matrices
$$A=\pmatrix{-{a'\dot q^R\over 2a} &
 -{\alpha'Ea'\over 2a}&{\bf O}\cr
           ~~&~~&~~\cr
	   -{\alpha'Ea'\over 2a} &
	   -{a'\dot q^R\over 2a}&
	   q^R\dot q^j\cr
	   ~~&~~&~~\cr
	   {\bf O} & {\dot q^i\over q^R}a&
	   {\dot q^R\over q^R}\delta^{ij}
	   +q^i\dot q^j\cr}~,\eqno(4)$$
$$B=\pmatrix{0&	-{\alpha'Ea''\dot q^R\over a}&{\bf O}\cr
             ~~&~~&~~\cr
             0& -S&{\bf O}\cr
             ~~&~~&~~\cr
	    {\bf O}&2\dot q^i\Bigl({a\over q^R}\dot{\Bigl)}
	      &\left({\alpha'L\over
	    (q^R)^2}\right)^2\delta^{ij}\cr}~,\eqno(5)$$
with $$S={a''\over 2a}\left[(q^R)^2+\alpha'^2E^2\right]-
\left({\alpha'L\over (q^R)^2}\right)^2a~,\eqno(6)$$
Here the dot stands for $\partial/\partial\tau$ and
$'=\partial/\partial q^R$. The solution for the string
coordinates is given by
the expansion
$$X^\mu=q^\mu+\eta^\mu+\xi^\mu+....~~,
{}~~\mu=0,1,....,D-1~.\eqno(7)$$
$q^\mu(\tau)$ being the center of mass
 coordinates (zeroth order solutions), which follow the
geodesics of the background space-time, i.e.
$$\dot q^0={\alpha'E\over a}~~,~~(\dot q^i)^2=
\left({\alpha'L\over (q^R)^2}\right)^2~~,
{}~~{(\dot q^R)^2\over\alpha'^2a}+{L^2\over(q^R)^2}+
m^2-{E^2\over a}=0~,\eqno(8)$$
We have identified the proper
time of the geodesic with
the world-sheet $\tau$-coordinate.

To study the equation of motion of the
first fluctuations, eq \(1),
it is convenient to apply a transformation
 to the vector $\eta^A$.
Let us propose
$$\eta_n=G\Xi_n~,\eqno(9)$$
where the matrix $G$ is chosen to
 eliminate the term in the first
derivative in eq \(1), i.e.
$$G=P\exp\{-\int^\tau A(\tau')d\tau'\}~,\eqno(10)$$
where $P$ is a constant normalizing matrix.
 Thus, eq \(1) transforms into
$$\ddot\Xi_n+G^{-1}(n^2+B-\dot A-A^2)G\Xi_n=0~,\eqno(11)$$
which is a Scr\"odinger-type equation with $\tau$
playing the role of the spatial coordinate.

\head{III - Head-on Collision}

Let us start with the simpler case of
a radial infall of a fundamental
string towards a black hole. In this
 case the transversal components of the
center of mass motion are zero, i.e.
 $q^i=\dot q^i=0$.

\noindent
a) {\it transverse coordinates:}
They  uncouple from the radial ones and
from each others giving rise to the
following equations
$$\ddot \eta^i_n+n^2\eta^i_n+2{\dot q^R\over
 q^R}\dot \eta^i_n=0~.\eqno(12)$$

By making the transformation
$$\Xi^i_n=q^R\eta^i_n~,\eqno(13)$$
eq \(12) yields
$$\ddot\Xi^i_n+(n^2-\ddot q^R/q^R)
\Xi^i_n=0~.\eqno(14)$$
And by use of the geodesics eqs \(8), we find
$$\ddot\Xi^i_n+\left[n^2+{(\alpha' m)^2\over 2}
{a'(q^R)\over q^R}\right]
\Xi^i_n=0~,\eqno(15)$$
where $'=\partial/\partial q^R$.

{}From the form of this equation we
can see that when the square bracket
is bigger than zero we will have the
 typical oscillatory motion of strings,
  but when the bracket reaches negative
  values, this equation
suggests the onset of instabilities.
 We can have imaginary frequencies only
in the case in which the local gravity, i.e.
 $a'/2$, is negative
 (repulsive effects). We know that for
 Schwarzschild black holes we have
$$a'_S(q^R)={(D-3)\over q^R}
\left({R_s\over q^R}\right)^{D-3}~,\eqno(16)$$
which is always bigger than
zero for $D>3$. Thus, we can
 conclude that the transversal
 modes are stable in the case of a
  radial infall towards a Schwarzschild black hole.

The case of a charged black hole gives the same
 result concerning the stability outside the
 event horizon, although, close to the singularity $(q^R\to0)$,
 we will have negative values of $a'$ (it is known
 that the Reissner-Nordstr\o m solution has a
 repulsive inner horizon\refto{rep}). In fact,
$$a'_{RN}(q^R)={(D-3)\over q^R}\left[
\left({R_s\over q^R}\right)^{D-3}-2
   \left({\tilde Q\over q^R}\right)^
   {2(D-3)}\right]\simeq-2{(D-3)\over q^R}
   \left({\tilde Q\over q^R}\right)^
   {2(D-3)}~.\eqno(17)$$
We see that the onset of the instability appears
within the event horizon. As the string falls towards
 the singularity $q^R\to0$, the first mode ($n=1$), begins to
 suffer instabilities, then the second mode too and so
  on. In this Reissner-Nordstr\o m case, the presence
  of a second inner horizon (usually denoted as $r_-$),
   implies $a'<0$ from somewhere in the region
   $r_-<q^R<r_+$ to $q^R\to0$, where $r_\pm=M\pm \sqrt{M^2-Q^2}$.

Including a positive cosmological constant will
again produce instabilities. In fact, for
the De Sitter space
$$a'_{DS}(q^R)=-{\Lambda\over3}q^R~.\eqno(18)$$
And thus, in particular for $M=0$ and $Q=0$,
 we recover the results \refto{vs87,sv90} about the onset of
 instabilities generated by the expansion of the
De Sitter universe, for large enough cosmological
constant, i.e.. $\alpha'm\Lambda/6>1$.
Obviously, anti-De Sitter universe ($\Lambda<0$),
will not generate instabilities. Thus, for the
complete case of a Reissner-Nordstr\o m black
hole immersed in a De Sitter universe, we will
have the possibility of instabilities far from
the black hole, where the cosmological solution
dominates and inside the black hole, where the
Reissner-Nordstr\o m solution dominates.

In the analysis above we have considered
$q^R$ as a parameter, but in general
it will be $\tau$-dependent. However, in
the approximation of first order fluctuations
we are considering, we can take $q^R$ as
 parametrizing the trajectory in an adiabatic
  approximation. We will see in the next section that
this is indeed the case, when we include in detail
 the time dependence of $q^R$.

\noindent
b) {\it  Radial coordinates}: Let us study now the
radial coordinates of the string.
 Here we have a
coupled set of equations for the first order
perturbations $\eta^0$ and $\eta^*$. To eliminate
 the first time derivative appearing in the equations
 of motion \(1), as we have seen, we can apply
  the matrix transformation $G$ given by expression
  \(10), which in our case, by use of the matrix
  $A(\tau)$, eq \(4), takes the following form
$$G={a^{1/2}\over\sqrt{\left| 1-\left({
\sqrt{E^2-m^2a}+E\over m}\right)^2\right|}}
\pmatrix {1&{\sqrt{E^2-m^2a}+E\over m}\cr
{\sqrt{E^2-m^2a}+E\over m}&1\cr}~, \eqno(19)$$
Here we have used
the geodesic equations \(8), as well as\refto{gr}
$$\int{da\over a\sqrt{E^2-am^2}}={1\over E}
\ln\left|{\sqrt{E^2-m^2a}-E\over
\sqrt{E^2-m^2a}+E}\right|~.\eqno(20)$$

It is not difficult to compute $\dot A$ and $A^2$
 from expression \(4). Thus, we have all the elements
  to write down explicitly the first order
  perturbations equations \(11)
$$\ddot\Xi_n+M\Xi_n=0~~,~~M=G^{-1}(n^2+B-
\dot A-A^2)G~,\eqno(23)$$
the matrix $M$ is given explicitly by
$$M={(\alpha')^2\over2(1-g^2)}\pmatrix{c-e+n^2&d-ge\cr
                                       d+e&c+ge+n^2\cr}
   ~,\eqno(24)$$
where
$$c=[(2E^2-am^2)(a'/a)^2+E^2(a''/a)](g^2-1)~,$$
$$d=[E\sqrt{E^2-am^2}(2(a'/a)^2+(a''/a))](g^2-1)
{}~,\eqno(26)$$
$$e=(a''/a)[2E\sqrt{E^2-am^2}-g(2E^2-am^2)$$
and
$$g={\sqrt{E^2-m^2a}+E\over m}~.\eqno(27)$$

Now, eq \(23) represents a set of two coupled
 equations for the components of the vector
$$\Xi_n=\pmatrix{\Xi_n^0\cr
                 \Xi_n^1\cr}~.\eqno(28)$$

In order to uncouple and better analyse eq \(23) we
 apply the following unitary transformation which
 diagonalizes the matrix $M$
$$\tilde M=TMT^{-1}~.$$
Thus, $\tilde M$ reads,
$$\tilde M=\pmatrix{\lambda_+&0\cr 0&
\lambda_-\cr}~,\eqno(31)$$
where the eigenvalues $\lambda_\pm$ are given by
$$\lambda_\pm={1\over2}[D+A\pm(4BC+(D-A)^2)^{1/2}]
+n^2~,\eqno(32)$$
and we have written for the matrix $M$ in eq \(24)
$$M\doteq\pmatrix{A&B\cr C&D\cr}~.$$
Thus, explicitly, the eigenvalues \(32) are given by
$$\lambda_\pm={(\alpha')^2\over4}\left[2(a'/a)^2
(2E^2-am^2)+m^2a''\pm\sqrt{
16E^2(E^2-am^2)(a'/a)^4+m^4(a'')^2}\right]+n^2~.
\eqno(34)$$
Now, this diagonalized expression allows us to
analyse the stability of the fluctuations around
the center of mass of the string. By looking for
 negative values of $\lambda_\pm$, we can scan
 the possibility of the onset of instabilities
 in the motion of strings. The analysis of eq \(34)
  for every value of $q^R$, can be made by replacing
   there the expression for $a(q^R)$ given by eq \(1').

It is convenient, in the Reissner-Nordstr\o m-(De Sitter)
 case, write expressions
 in terms of the following variable
$$x^{-1}\doteq\left({R_s\over q^R}\right)^{D-3}~~,
{}~~\beta\doteq\left({\tilde Q\over R_s}\right)^
{2(D-3)}~.\eqno(36)$$
Thus, eq\(1') reads,
$$a(x)=1-{1\over x}+{\beta\over x^2}+\bigg(-
{\Lambda \over3}(q^R)^2\bigg)~.\eqno(35)$$

It can be seen that for $m=0$
$$\lambda_+=n^2+2\alpha'^2\left({a'\over a}\right)^2E^2~,$$
$$\lambda_-=n^2~,\eqno(35')$$
that is, for $m=0$ the evolution is {\it stable} in
the three cases: Schwarzschild, Reissner -
Nordstr\o m and De Sitter spacetimes. $\lambda_+$
(and then $\Sigma^0$) is always stable (this is so, even for
$m\not=0$) and $\lambda_-$ is as in flat spacetime; the
first order fluctuations for the massless string always oscillate.

\noindent
c) {\it Analysis}: To simplify the analysis,
 we will see what happens in three important
 regions of the black hole space: far away
 from the black hole, close to the event
 horizon and approaching the singularity in
 the Schwarzschild, Reissner-Nordstr\o m and
  De Sitter spacetimes.

i) {$q^R\to\infty$}: In this case the terms
 proportional to $a''$ dominate over those
 proportional to $(a'/a)^2$ and thus we have
$$\lim_{q^R\to\infty}{\lambda_\pm^\infty}=
{(\alpha')^2\over4}\left[2(a'/a)^2(2E^2-am^2)
+m^2(a''\pm\left| a''\right|)\right]+
n^2~.\eqno(39)$$
Thus, from expressions \(36)-\(35), we can
conclude that the modes with coefficient
$ \lambda_+$, being definite positives,
are stable in the
Schwarzschild and Reissner-Nordstr\o m
black holes, i.e.
$$\lim_{q^R\to\infty}{\lambda_+^{BH}}=
{(\alpha')^2\over2}(D-3)^2{(2E^2-m^2)
\over (q^R)^2x^2}+n^2~,\eqno(40)$$
 as well as in the De Sitter space time, i.e.
 $$\lim_{q^R\to\infty}{\lambda_+^{DS}}=
 2(\alpha')^2\left[{m^2\Lambda\over3}+
 {(2E^2-m^2)\over (q^R)^2}\right]+
 n^2~.\eqno(41)$$

 The other eigenvalue,$ \lambda_-$, however,
  would indicate the emergence of an instability
  for not so large values of $q^R$. In fact, in
  this case we have for the Schwarzschild
  and Reissner-Nordstr\o m black holes,
$$\lim_{q^R\to\infty}{\lambda_-^{BH}}={(\alpha')
^2\over2}(D-3)(D-2){m^2
\over (q^R)^2x}+n^2~.\eqno(42)$$
 And for the De Sitter case,
$$\lim_{q^R\to\infty}{\lambda_-^{DS}}=-{(\alpha')
^2\over3}m^2\Lambda+n^2~.\eqno(43)$$

We observe that in $\lambda_-$ appears a fundamental
 difference with respect to
$\lambda_+$: The term proportional to $m^2$,
 is here negative. This allows the possibility of the
onset of instabilities for values of $m^2$
large enough to
unstabilize successively the modes $n=1, 2, ...$
and so on. In the case of the Schwarzschild or
Reissner-Nordstr\o m black holes, for large
values of $q^R$, we have stability, even for
the $\lambda_-$ modes, as one would have expected
 due to the asymptotic flatness of the spacetime.

ii) $q^R\to q^R_{Horizon}$: In this case the
terms proportional to $(a'/a)^2$ dominate over
 those proportional to $a''$ and we have
$$\lim_{q^R\to q^R_H}{\lambda_\pm^H}=n^2+{(\alpha')
^2\over4}
\left[2(a'/a)^2(2E^2-am^2\pm2E\sqrt{E^2-am^2})+
m^2a''\right]~.\eqno(43')$$
Again, we will see that the modes $\lambda_+$
yield stable fluctuations around the center of
 mass. In fact, for black holes and De Sitter cases
$$\lim_{q^R\to q^R_H}{\lambda_+^H}=n^2+(\alpha')^
2(a'/a)^2(2E^2-am^2)~,
\eqno(44)$$
which is always a positive quantity.

The other mode $\lambda_-$ carries again the
 possibility for the emergence of instabilities.
For the Schwarzschild black hole we have
$$\lim_{q^R\to q^R_H}{\lambda_-^{Sch}}=
n^2+{(\alpha')^2(D-3)\over4}{m^2\over R_s^2}
\left[2-D+{(D-3)\over2}{m^2\over E^2}\right]=\eqno(49)$$
$$=n^2-{(\alpha')^2m^2\over 2R_s^2}\left[1-{m^2\over 4E^2}\right]
{}~~D=4~.$$

As the term between brackets is always negative
 for $D\geq3$, this indicates that we have
 the possibility of an unstable regime (depending
 on the value of ${m^2/R_s^2}$) for the first
 excited modes. The instability condition can be written as
$$n<{S\over R_s}\left[D-2-{(D-3)\over2}{m^2\over E^2}\right]
^{1/2}\eqno(49')$$
where the quantity between brackets is always positive and
$S=\alpha'm\sqrt{D-3}$ is a measure of the string
size. Thus, if the string is larger than the horizon radius,
it become unstable. This is similar to the instability criterion
in De Sitter space, i.e. $n<\alpha'm/r_H$, $r_H=H^{-1}$
being the horizon radius.

This is not the case, however, for the extreme
Reissner-Nordstr\o m black hole. In fact,
$$\lim_{q^R\to q^R_H}{\lambda_-^{ERN}}=(\alpha')
^22^{D+1\over 3-D}(D-3)^2{m^2\over R_s^2}+n^2~.
\eqno(50)$$
which is always a positive quantity and does
 not develop instabilities at first perturbative
 order. This shows that there is a critical
 value of the electric charge of the Reissner
 -Nordstr\o m black hole above which the
 string on the horizon pass from unstable to
 stable regime. This critical value can be
  found by making vanish $\lambda_-^H$ in eq \(43').

For the De Sitter space
$$\lim_{q^R\to q^R_H}{\lambda_-^{DS}}={(\alpha')
^2\over6}m^2\Lambda\left(
{m^2\over E^2}-1\right)+n^2~.\eqno(51)$$
As the quantity between the parenthesis is always
 negative for the radial orbits we are studying,
  we could have the onset of instabilities, again
  this depending on having large enough values of
   $m^2\Lambda$.

iii) {$q^R\to0$}: This limit gives the approach
to the singularity. Here, terms proportional to
$a''$ dominate over those proportional to $(a'/a)^2$.
 We have, then, in this limit
$$\lim_{q^R\to0}{\lambda_\pm^0}={(\alpha')^2\over4}
\left[2(a'/a)^2(2E^2-am^2)+m^2(a''\pm\left| a''
\right|)\right]+n^2~.\eqno(52)$$
We will see, again, that as $q^R\to0$, $\lambda_+^0$ remains
positive, thus giving stable first order fluctuations.

In fact, for Schwarzschild black holes
$$\lim_{q^R\to0}{\lambda_+^{Sch}}={(\alpha')
^2\over2}{(D-3)^2m^2
\over (q^R)^2x}+n^2~,\eqno(53)$$
which is always positive. Thus, producing stable
first order fluctuations.

The same happens for the Reissner-Nordstr\o m
black holes,
$$\lim_{q^R\to0}{\lambda_+^{RN}}={(\alpha')^2
\over2}{(D-3)(4D-10)\beta m^2
\over (q^R)^2x^2}+n^2~.\eqno(54)$$
And also for the case of the De Sitter space,
$$\lim_{q^R\to0}{\lambda_+^{DS}}={2\over9}
(\alpha')^2\Lambda^2(q^R)^2(2E^2-m^2)
+n^2~,\eqno(55)$$
the first order oscillations will be bounded.

Again, the situation changes when we study the mode
 $\lambda_-$. For the Schwarzschild black
 hole we have
$$\lim_{q^R\to0}{\lambda_-^{Sch}}=-{(\alpha')
^2\over2}{(D-3)(D-2)m^2
\over (q^R)^2x}+n^2~,\eqno(56)$$
which suggest the onset of instabilities.
We remark here, the similarity between this
expression and eq \(42) for the case $q^R\to\infty$.

The Reissner-Nordstr\o m case yields also
 the possibility of instabilities,
$$\lim_{q^R\to0}{\lambda_-^{RN}}=
-2{(\alpha')^2}{(D-3)^2\beta m^2
\over (q^R)^2x^2}+n^2~.\eqno(57)$$
And finally, the De Sitter space-time
gives an unstable solution too,
$$\lim_{q^R\to0}{\lambda_-^{DS}}=
-{1\over3}(\alpha')^2\Lambda m^2
+n^2~.\eqno(58)$$
We remark again the similarity with
the result eq\(43), valid for $q^R\to\infty$.
\bigskip

Several remarks worth doing here. First,
we observe from table 1 that
 the longitudinal modes $\lambda_+$ always
  give stable fluctuations, while the
  $\lambda_-$ modes almost always suggest the
  existence of instabilities. By choosing an
  appropriate gauge in the $\Xi^0$ and $\Xi^1$
  coordinates, we can interpret the mode $\lambda_+$
   as the temporal coordinate and the mode proportional
    to $\lambda_-$ as the radial coordinate.
    As for the
cosmological backgrounds\refto{sv90},
 the string time coordinate is well behaved.

We have already remarked that in the Schwarzschild
 and De Sitter spacetimes the modes $\lambda_-$
 and $\Xi^i$ are quite similar in the
  limits $q^R\to0$ and $q^R\to\infty$. This
  property has also been found in cosmological
    backgrounds\refto{v91}.

We have studied separately the cases of black
holes and De Sitter space. We have done so for
 the sake of simplicity and for the importance
 of De Sitter space in itself, but it is
 straightforward to observe the regimes in which
  one case predominates over the other when we
  study a black hole embedded in the De Sitter
  spacetime: At large distances from the hole,
  the cosmological term dominates, while not
  far from the black hole their contribution
  can be neglected.

Here, as remarked in references [\cite{vs87,sv90}],
in order to discuss the constraints imposed to the
string equations of motion, one has to go
to second order string fluctuations. To first
 order, constraints are satisfied consistently
 with the equations of motion only for stable modes.

It is important to stress that the perturbative
analysis of the equations of motion we have done,
 is strictly valid in the stable regimes
 and allows to discover the presence of
 instabilities.
  In the unstable cases, one has to describe
  the unstable non linear regime, non-perturbatively.
  Asymptotic  solutions describing the highly
  unstable string regime are under study
    by the present authors and will be published
   elsewhere\refto{LS93}.

\head{IV - Evolution with time}

In the last section we have considered $q^R$,
the coordinate of the center of mass as a parameter.
 This allowed us to analyse the
  stability of the fluctuations of fundamental
  strings. Now, in order
   to see the evolution with the
  proper time $\tau$, let us consider $q^R(\tau)$
   and integrate the resulting differential
   equation for $\Xi(\tau)$,
$$\ddot\Xi_n^\mu+[n^2+\lambda^\mu]\Xi_n^\mu=0~,
\eqno(59)$$
where
$$\Xi_n^\mu=(\Xi_n^+, \Xi_n^-, \Xi_n^i)~~;
{}~~\lambda^\mu=(\lambda^+, \lambda^-,
\lambda^i)~,\eqno(60)$$
and
$$\lambda^\pm=\lambda_\pm-n^2~~,~~\lambda^i
={(\alpha' m)^2\over 2}
{a'(q^R)\over q^R}~.\eqno(70)$$
We can obtain the time dependence of
$q^R$ from the geodesics equations, \(8). Thus,
$$\alpha'(\tau-\tau_0)=\int^{q^R}_{q^R_0}
{dq^R\over\sqrt{E^2-m^2a(q^R)}}~. \eqno(71)$$

Let us now study the different curved
backgrounds we are interested in.

\noindent
a) De Sitter space

This case is the simplest for analysing,
 due to the constancy of $\lambda^i$,
$$\lambda^i_{DS}=-{(\alpha'm)^2\over6}
\Lambda~.\eqno(72)$$
Thus, eq \(59) for the transverse modes
can be easily solved in
 terms of exponentials, i.e.
$$\Xi_n^i(\tau)=\exp\left\{\pm i\left[n^2
-{(\alpha'm)^2\over6}\Lambda\right]^{1/2}
\tau\right\}~.\eqno(73)$$
We see that for a cosmological
constant positive and bigger than $6/(\alpha'm)$,
 the first mode $\Xi_1^i(\tau)$ will begin to grow
  exponentially with time. For even  bigger values
 of $\Lambda$ further modes can be excited. Negative
 values of $\Lambda$ (anti-De Sitter space) gives
  bounded fluctuations.

The analysis for the longitudinal modes is
quite more complicated, because of the time
dependence of $\lambda^\pm_{DS}$. Thus, we will
 consider the two asymptotic regions  $q^R\to0$
 and $q^R\to\infty$.

$q^R(\tau)$ can be obtained explicitly by
 integrating\refto{gr} eq \(71),
$$\alpha'(\tau-\tau_0)=\cases{{1\over m}
\sqrt{3\over\Lambda}{\rm arcsinh}\left(
{m\over \sqrt{E^2-m^2}}\sqrt{\Lambda\over3}
q^R\right)~~,~~~~~~~{\rm for~ \Lambda>0}\cr
-{1\over m}\sqrt{3\over-\Lambda}{\rm arcsin}
\left({m\over \sqrt{E^2-m^2}}\sqrt{-\Lambda
\over3}q^R\right)~~,~~~{\rm for ~\Lambda<0.
}\cr} \eqno(74)$$
where $\tau_0$ is the proper time of arrival to $q^R=0$.

In the limit $q^R\to\infty$, the eigenvalues
 $\lambda^\pm_{DS}$
 are given by eqs\(41) and \(43), thus,
 having constant values,
$$\lambda^+_{DS}\to{2(\alpha'm)^2\over3}
\Lambda~,$$
$$\lambda^-_{DS}\to-{(\alpha'm)^2\over3}
\Lambda~.\eqno(75)$$
With these constant values, eq \(59) can
be easily solved for the $\Xi^\pm$ coordinates,
 again in terms of exponentials. The same
 analysis and the same
 condition of instability as
 for the transversal modes apply here for
 $\lambda^-$.  This is so because we have
  negative values of $\lambda^-$ for
  $\Lambda>0$. For $\lambda^+$, instead,
   we have bounded solutions.

In the limit $q^R\to0$ we must use expressions
\(55) and \(58). Again $\lambda^-$ gives a
constant negative value which indicates the
emergence of instability. $\Xi^-$ behaves like
$$\Xi^-_{n, DS}\to\exp\left\{\pm i\left[n^2-
{(\alpha'm)^2\over3} \Lambda\right]^{1/2}
\tau\right\}~.\eqno(76)$$
On the other hand, from eqs \(55) and \(74),
 we find that
$$\lambda^+_{DS}\simeq {2\over9}(\alpha')^4
\Lambda^2(E^2-m^2)(\tau-\tau_0)^2 \doteq
C(\tau-\tau_0)^2~,\eqno(77)$$
which inserted in eq \(59) gives a regular
 behavior for $\Xi^+_n$ in terms of a power
  series of $(\tau-\tau_0)$, i.e.
$$\Xi^+_{n, DS}(\tau)=\sum_{j=0}^\infty K_j
(\tau-\tau_0)^j~.\eqno(78)$$
Here $K_j$ can be found from the recursive
formula
$$(j+2)(j+1)K_{j+2}+n^2K_j+CK_{j-2}=0~~;
{}~~j=0,1,2,.....~~,~~K_j=0~~{\rm for~ j<0}
{}~.\eqno(79)$$
and $K_0=\Xi_0$ and $K_1=p^+$ are the
 initial data.

\noindent
b) Schwarzschild black hole

This case is somewhat more complicated algebraically
than the De Sitter one, but it can be analysed
in the two interesting asymptotic regions
 $q^R\to0$ and $q^R\to\infty$.

When $q^R\to0$, the geodesic path can be
 approximated by
$$q^R_S(\tau)\to\left[{(D-1)\over2}
\alpha'R_s^{(D-3)/2}m(\tau-\tau_0)\right]
^{2/(D-1)}~,\eqno(80)$$
$$q^0_S(\tau)\sim(\tau-\tau_0)^{2(D-1)/(D+1)}\to0~,$$
where $\tau_0$ is the proper time of arrival to the singularity at $q^R=0$.

Replacing this path,$q^R_S(\tau)$, into eq \(59), we have for
 the transversal coordinates
$$\ddot\Xi_n^i+\left[n^2+{2(D-3)\over(D-1)
^2}(\tau-\tau_0)^{-2}\right]\Xi_n^i=0~,
\eqno(81)$$
for which we find the power-law solution,
$$\Xi_{n, Sch}^i(\tau)\sim \left[n(\tau-
\tau_0)\right]^P~~;~~P^i_{Sch}={1\over2}
\pm\sqrt{{1\over4}-{2(D-3)
\over(D-1)^2}}~.\eqno(82)$$
For $D\geq4$, $P$ is always real and
positive. Thus, in the regime studied,
 $(\tau-\tau_0)$ small, we have regular
  and bounded solutions $\Sigma^i$. And so confirms
   that there is not instabilities in
    this region.

{}From expressions \(53) and \(56), we observe
that $\lambda^\pm$ produce equations analogous
 to eq \(81) and power-like solutions as
 in expression \(82). Only the value of
 $P$ changes,
$$P^+_{Sch}={1\over2}\pm\sqrt{{1\over4}-
2{\left(D-3
\over D-1\right)^2}}~.\eqno(83)$$
This gives complex solutions for $D>5$,
 but anyway bounded as $(\tau-\tau_0)\to0$.

On the other hand,
$$P^-_{Sch}={1\over2}\pm\sqrt{{1\over4}+
{2(D-3)(D-2)
\over(D-1)^2}}~,\eqno(84)$$
gives a solution, which with the minus sign
in front
of the square root is unbounded
as $(\tau-\tau_0)\to0$. Thus, being consistent
 with the results of the last section (see
 comments made after eq \(56)
 about the possibility of having
 instabilities in this longitudinal mode
  of the string).

The same kind of analysis can be made
 in the region $q^R\to\infty$. The geodesic
 equations in this case yields
$$q^R_S(\tau)\to\alpha'\sqrt{E^2-m^2}
(\tau-\tau_0)~.\eqno(85)$$
Thus, the equation for the transversal
modes is
$$\ddot\Xi_n^i+\left[n^2+D^i(\tau-\tau_0)^
{1-D}\right]\Xi_n^i=0~,\eqno(86)$$
where
$$D^i={(D-3)\over2}{m^2\over(E^2-m^2)}
\left({R_s\over\alpha'\sqrt{E^2-m^2}}
\right)^{D-3}~.$$

While for the longitudinal modes we have,
$$\ddot\Xi_n^-+\left[n^2+D^-(\tau-\tau_0)^
{1-D}\right]\Xi_n^-=0~,\eqno(87)$$
with
$$D^-=-{(D-3)\over2}{(D-2)m^2\over(
E^2-m^2)}\left({R_s\over\alpha'
\sqrt{E^2-m^2}}\right)^{D-3}~.$$
and
$$\ddot\Xi_n^++\left[n^2+D^+(\tau-\tau_0)
^{4-2D}\right]\Xi_n^+=0~,\eqno(88)$$
where
$$D^+={(D-3)^2\over2}(\alpha')^{6-2D}
{(2E^2-m^2)\over(E^2-m^2)^{D-2}}
R_s^{2(D-3)}~.$$

Eqs \(86)-\(88) can be solved in terms
of Laurent series (negative power series
of $(\tau-\tau_0)$), that in the limit $
(\tau-\tau_0)\to\infty$ makes them convergent
 solutions. Only the negative value in $D^-$
 suggest that for $(\tau-\tau_0)$  not so large,
  some kind of instabilities could occur.

\noindent
c) Reissner-Nordstr\o m black hole

For charged black holes, in the limit of
$q^R\to0$, the gravitational effect of the
charge overwhelms that of the mass. In that
 limit, from eqs \(1') and \(71), we obtain
$$q^R(\tau)\to\left[\alpha'(D-2)m(\tau-\tau_0)
\tilde Q^{D-3}\right]^{1/(D-2)}~. \eqno(89)$$
Plugging this expression into eq \(59) and
 using \(17) and \(70), we obtain
$$\ddot\Xi_n^i+\left[n^2-{(D-3)\over(D-2)^2}
(\tau-\tau_0)^{-2}\right]\Xi_n^i=0~,\eqno(90)$$
this is as eq \(81) for the transversal
oscillations in a Schwarzschild black hole,
 but the coefficient in front of $(\tau-\tau_0)
 ^{-2}$ has now, a negative value. This makes
 that the solution
$$\Xi_{n, RN}^i(\tau)\sim \left[n(\tau-\tau_0)
\right]^P~~;~~P^i_{RN}={1\over2}\pm\sqrt{
{1\over4}+{(D-3)
\over(D-1)^2}}~,\eqno(91)$$
allows a solution (that with minus sign
 in front of the
square root), which is unbounded as $(\tau-\tau_0)
\to0$, and indicates an instability
as seen in the analysis of the previous section.

{}From the eqs \(54) and \(57) for $\lambda^\pm$
 we see that the longitudinal coordinates of
 the string in the Reissner-Nordstr\o m background,
  as it falls towards the singularity at $r=0$,
   will behave like a power $P_\pm$ of the
    proper time,
$$P^+_{RN}={1\over2}\pm\sqrt{{1\over4}-
{(D-3)(4D-10)\beta
\over2(D-1)^2}}~.\eqno(92)$$

For $D\geq4$, this exponent becomes complex
 (for big enough $\beta$), but still produces
 the fluctuations $\Xi^+$ to vanish as
  $(\tau-\tau_0)\to0$.

The case for $\Xi^-$ is different because
$$P^-_{RN}={1\over2}\pm\sqrt{{1\over4}+
2{(D-3)^2\beta
\over(D-1)^2}}~,\eqno(93)$$
allows again one solution that is unbounded
 as $(\tau-\tau_0)\to0$. Thus confirming our
  analysis of the last section
   for the Reissner - Nordstr\o m
   black hole in this region.

The time evolution in the intermediate region
 between the two asymptotic ones studied, can be
 analysed from the results obtained in the
  last section. In fact, on the event
  horizon we have the following behavior,
$$\Xi^\mu_{n, H}(\tau)\sim\exp\{\pm i\left[
n^2+\lambda^\mu_H\right]^{1/2}(\tau-\tau_0)\}
{}~,\eqno(94)$$
where $\lambda^\mu_H$ can be found
from eqs \(44)-\(51). $\lambda^+_H$ is always positive
while  $\lambda^-_H$ and $\lambda^i_H$ can take negative
values depending on the string and black hole parameters.
We can thus draw
the same conclusions than before about
the conditions of stability: Near the horizon, the string
behavior is stable and oscillatory
for the time coordinate $\Sigma^+$ and for the higher $n$ modes
of the radial and transversal components; the lower $\Sigma^-$
and $\Sigma^i$ modes being, however, unstables.

In summary, the time-dependent analysis
 confirms completely the results about
 stability and about the onset of instabilities
obtained in the last section on the grounds
 of  an adiabatic study.

\head{V - Non-colinear collision and discussion}

It is interesting to investigate how
changes the picture when the infalling
string has an orbit with non zero impact parameter.

It is simple to analyse the first order
fluctuations in the transversal coordinates.
 In fact, for $i>2$, the matrices $A$ and $B$
  given by eqs \(4) and \(5) are diagonal,

$$A^{ij}={\dot q^R\over q^R}\delta^{ij}~~;
{}~~B^{ij}=\left({\alpha'L\over (q^R)^2}\right)
^2\delta^{ij}~,\eqno(102)$$
then the equations for the first order
fluctuations are
$$\ddot\Xi^i_n+(n^2+\left({\alpha'L\over
 (q^R)^2}\right)^2-\ddot q^R/q^R)\Xi^i
 _n=0~.\eqno(103)$$
This is the generalization of eq \(14)
 for $L\not=0$.

By use of the geodesic equations \(8)
 we can rewrite eq \(103) as
$$\ddot\Xi^i_n+\left[n^2+{(\alpha')^2
\over 2}\left({a'(q^R)m^2\over q^R}+
2\left({L\over (q^R)^2}\right)^2\left(
1-a+{1\over2}a'q^R\right)\right)\right]
\Xi^i_n=0~.\eqno(104)$$
Thus we are able to analyse now the particular
 cases we are interested in:

\noindent
a) Schwarzschild black hole:

Plugging eqs \(1') and \(16) into \(104)
 we obtain
$$\lambda^i_{Sch}={(\alpha')^2\over 2}R_s^
{D-3}(q^R)^{1-D}\left[(D-3)m^2+(D-1)\left(
{L\over q^R}\right)^2\right]~.\eqno(105)$$
We observe that for $D\geq4$, $\lambda^i_
{Sch}>0$. Thus producing stable first order
 fluctuations. In fact, when we study the
 time evolution we obtain
$$\lim_{q^R\to0}{\lambda^i_{Sch}}=
{2(D-1)\over(D+1)^2}(\tau-\tau_0)^{-2}
{}~,\eqno(106)$$
where we have integrated the expression \(8),
$$\alpha'(\tau-\tau_0)=\int^{q^R}_{q^R_0}
{dq^R\over\sqrt{E^2-m^2a(q^R)-a\left({L\over
 q^R}\right)^2}}~, \eqno(107')$$
in order to obtain the behavior
$$q^R_S(\tau)\to\left[{(D+1)\over2}\alpha'R_s
^{(D-3)/2}L(\tau-\tau_0)\right]^{2/(D+1)}.
\eqno(107)$$
Eq \(105) has, then, a power-like solution
in the limit $q^R\to0$,
$$\Xi_{n, Sch}^i(\tau)\sim \left[n(\tau-\tau_0)
\right]^P~~;~~P^i_{Sch}={1\over2}\pm
\sqrt{{1\over4}-{2(D-1)
\over(D+1)^2}}~.\eqno(108)$$
That vanishes for $(\tau-\tau_0)\to0$.

\noindent
b) De Sitter spacetime

This case is very interesting because replacing
 expressions \(1') and \(18) into the first
 order fluctuations equations with $L\not=0$,
  eq \(104), we obtain that the $L$
  dependence disappears. Giving, in fact,
   the equation
$$\ddot\Xi^i_n+\left[n^2-{(\alpha'm)^
2\over6}\Lambda\right]
\Xi^i_n=0~.\eqno(109)$$
The solution to this equation are exponentials
 in the proper time, $\tau$,
  and are given by eq \(73).
  The same conclusions about the possibility
  of instabilities given after eq \(73)
  can be drawn here.

The fact that the solutions
should be $L$-independent,
could have been guessed from the symmetries
 of the De Sitter space. This metric has
  not preferred point to refer the angular
 moment to as in the black hole cases (there
  is not singularity at $r=0$). This allows
  us to say that the components $Xi^-$,
   $Xi^+$ and $Xi^i$ will behave as the
   ones already studied for the case $L=0$.

\noindent
c) Reissner-Nordstr\o m black hole

{}From the metric coefficients \(1') and its
first derivative we find that eq \(104) reads
$$\ddot\Xi^i_n+\left\{n^2+{(\alpha')^2\over2}
\left[{(D-3) m^2
\over (q^R)^2x}(1-{2\beta\over x})+{L^2\over
x(q^R)^4}\left((D-1)-{2\beta\over x}(D-2)
 \right)\right]\right\}\Xi^i_n=0~.\eqno(110)$$
We observe that as $q^R\to\infty$, the term
proportional to $L^2$ vanish faster than the
other terms. Thus, in this limit we recover
the $L=0$ results.

As we approach to the black hole, and arrive
to the horizon, we have for the Schwarz\-schild
 black hole
$$\lim_{q^R\to q^R_H}{\lambda^i_{H,Sch}}=
{(\alpha')^2\over2}\left[
{(D-3)m^2\over R_s^2}+(D-1){L^2\over R_s^4}
\right]~,\eqno(111)$$
which is definite positive, and does not produce
 instabilities.

For the extreme Reissner-Nordstr\o m black
hole we have,
$$\lim_{q^R\to q^R_H}{\lambda^i_{H,ERN}}=
(\alpha')^22^{4/(D-3)}{L^2\over R_s^4}
{}~.\eqno(112)$$
Again, it is always positive. Thus, we see
 that the instabilities do not appear yet.
  The time-dependence of the solutions
  close to the horizon will be oscillatory
   with the squared frequency given by
   ${\lambda^i_{H,ERN}}$.

However, the picture changes when we go
closer to the singularity. For the Reissner-Nordstr\o m
black hole, when $q^R\to0$, we have
$$\lim_{q^R\to0}{\lambda^i_{0,RN}}=
-(\alpha')^2
{(D-2)\beta\over x^2}{L^2\over (q^R)^4}
{}~.\eqno(113)$$
As this squared frequency takes negative
values  allows the possibility that
instabilities develop in
the string transversal coordinates.
In order to find the time dependence in the
coordinates we integrate first the
center of mass motion, eq \(8). Then,
as $q^R\to0$, we obtain
 $$\alpha'(\tau-\tau_0)\sim{x\over L\sqrt{
 \beta}}{(q^R)^2\over(D-1)}~.\eqno(114)$$
Plugging this expression into eq \(113),
 we have
$$\lim_{q^R\to0}{\lambda^i_{0,RN}}\to-
{(D-2)\over(D-1)^2}(\tau-\tau_0)^{-2}
{}~. \eqno(115)$$
And the solution of the first order
 fluctuations is again a power-like one,
$$\Xi_{n, RN}^i(\tau)\sim \left[n(\tau-
\tau_0)\right]^P~~;~~P^i_{RN}={1\over2}
\pm\sqrt{{1\over4}+{(D-2)
\over(D-1)^2}}~.\eqno(116)$$
We have here that the solution with minus
sign in front of the square root produces
an unbounded solution as $(\tau-\tau_0)\to0$,
 thus, suggesting the existence of
 instabilities.

It is worth to remark that the solutions
\(116) and \(108) for the time dependence
 of the transversal coordinates for the
 Reissner-Nordstr\o m and Schwarzschild
 black holes respectively, are independent
 of $L$. They are, although, different from
 those of the case $L=0$ (eqs \(91) and \(83)).
This is so because even if the $L$
dependence cancels out from the final
equations \(116) and (108), the approach
to the singularity, $q^R\to0$, is
different if $L\not=0$, thus producing
different final coefficients.

Another interesting feature of the equations
for the transversal first order modes, is
that for the black hole cases (Schwarzschild
or Reissner-Nordstr\o m; orbit of the
string center of mass with or without
angular momentum), the time dependence
of $\lambda^i$ appears to be $(\tau-\tau_0)^{-2}$
as $q^R\to0$. The behavior of $\lambda^i$
as a function of $q^R$ is different for each case,
but the $\tau$ - dependence of the orbit in each case
exactly compensates for such difference.

Thus, for the linearized string fluctuations,
the approach to the black hole singularity
corresponds to the case $\beta=2$ of the motion of a particle
in a potential $\gamma(\tau-\tau_0)^{-\beta}$. This is like
the case of strings in singular
plane - wave backgrounds\refto{vs91,vms}. In fact, the linearized
first order string fluctuations produce a one-
dimensional Schr\"odinger equation, with
$\tau$ playing the role of a spatial coordinate.
The potential term in eq \(104), can be
written fully $\tau$ - dependent, as we
have seen, by plugging into it the center
of mass trajectory, $q^R(\tau)$.(In the case of gravitational
plane waves, the spacetime is simpler than  in the black hole
one, and the exact full string equations become linear).

The solution of eq \(104) with a potential
proportional to $\gamma(\tau-\tau_0)^{-2}$
can be given in terms of Bessel functions,
$$\Xi_{n}^i(\tau)=\sqrt{(\tau-\tau_0)}\left
\{V_{n}^iJ_\nu[n(\tau-\tau_0)]+
W_{n}^iJ_{-\nu}[n(\tau-\tau_0)]\right\}
{}~.\eqno(117)$$
Where $V_{n}^i$ and $W_{n}^i$ are arbitrary
constants coefficients and
$$\nu=\sqrt{{1\over4}-\gamma}~~~~{\rm i.e.}
{}~,~~\nu=P^\mu-1/2~,\eqno(118)$$
Where $P^\mu(D,\beta)$'s are defined in Section IV.

For $\gamma<0$ we have
Bessel functions (those with negative index)
with a divergent behavior as $(\tau-\tau_0)\to0$,
indicating the existence of
string instabilities.
We would also to stress that for black
holes, what determinates the possibility
of instabilities is not the type of singularity
(Reissner-Nordstr\o m or Schwarzschild),
nor how it is approached ($L=0$ or $L\not=0$),
because we have seen that the time dependence
of the potential is always $(\tau-\tau_0)^{-2}$,
 but the sign of the coefficient $\gamma$ in
front of it, i.e. the attractive character of the potential
$(\tau-\tau_0)^{-2}$. Thus, we can
conclude that whenever we have big enough
repulsive effects in a gravitational background,
instabilities in the
propagation of strings on the
spacetime background will appear.
The coefficient $\gamma$ is given by
$$\gamma_{Sch}=\cases{{2(D-3)\over(D-1)^2}~~,~~L=0\cr
\cr
                      {2(D-1)\over(D+1)^2}~~,~~L\not=0\cr}$$
for the Schwarzschild black hole, and
$$\gamma_{RN}=\cases{-{2(D-3)\over(D-2)^2}~~,~~L=0\cr
\cr
                      {-2(D-2)\over(D-1)^2}~~,~~L\not=0\cr}$$
for the Reissner-Nordstr\o m black hole.

Near the spacetime singularity, the dependence on the $D$ spacetime
dimensions is concentrated in $\gamma$. Notice the attractive singular
character of the potential $\gamma(\tau-\tau_0)^{-2}$, for the
Reissner-Nordstr\o m black hole, in agreement with the singular behavior
of the string near $q^R=0$; while for the Schwarzschild black hole
$\gamma$ is positive, and the string solutions $\Sigma^i$ are well
behaved there.

The approach to the black hole singularity is better analysed in
terms of the Kruskal coordinates $(u_k, v_k)$
$$u_k=e^{Ku_{Sch}}~~~,~~~v_k=e^{Kv_{Sch}}$$
$u$ and $v$ being null coordinates, and $K$ the surface gravity of the
black hole $(K=(D-3)/(2R_s)$ for Schwarzschild). From eqs \(82) - \(84)
we have $(C^\pm(\sigma)$ being coefficients determined by the initial
state of the string)
$$u_k(\sigma,\tau)=\exp{K\left[C^+(\sigma)(\tau-\tau_0)^{P^+}-
C^-(\sigma)(\tau-\tau_0)^{-|P^-|}\right]}$$
$$v_k(\sigma,\tau)=\exp{K\left[C^+(\sigma)(\tau-\tau_0)^{P^+}+
C^-(\sigma)(\tau-\tau_0)^{-|P^-|}\right]}\eqno(u1)$$
and thus, $u_kv_k\to1$ for $(\tau-\tau_0)\to0$.
 That is, for $(\tau-\tau_0)\to0$, the
string approaches the spacetime singularity $u_kv_k=1$, and it is
trapped by it. The proper spatial length element of the string at
fixed $(\tau-\tau_0)\to0$, between $(\sigma,\tau)$ and $(\sigma+d\sigma,\tau)$,
stretches infinitely as
$$ds^2_{(\tau-\tau_0)\to0}\to \left(C^{-}(\sigma)\right)'^2d\sigma^2
(\tau-\tau_0)^{-(D-1)|P^-|}~,\eqno(u2)$$
where $P^-$ is given by eq \(84). Here, $\tau_0$ is the (finite) proper
falling time of the string into the black hole singularity.

The fact that the angular coordinates $\Sigma^i$ become un\-bounded in the
Reissner - Nordstr\o m case, means that the string makes infinite turns
around the spacetime singularity and remains trapped by it.

The same conclusions can be drawn for
the quantum propagation of
strings. The $\tau$ dependence is the same
because this is formally described
by the same Schr\"odinger equation
with a potential $\gamma(\tau-\tau_0)^{-2}$,
the coefficients of the solutions being quantum operators instead of
C - numbers. The $\tau$ evolution of the string near the black hole
singularity is fully determined by the spacetime geometry, while the
$\sigma$ - dependence (contained in the overall coefficients) is fixed
by the state of the string.

For the modes $\Xi^-$, $\Xi^+$ and $\Xi^i$
we can conclude that they should behave as
in the case $L=0$ far from the black hole,
where the influence of the angular momentum
vanishes. Then, $\Xi^+$ and $\Xi^i$ will
oscillate with bounded amplitude outside the horizon while
$\Xi^-$ will present an unbounded behavior.
The approach to the singularity with
$L\not=0$ should not change qualitatively
from the picture for $L=0$.
The analysis can be also made in terms of
the geodesic orbit followed by the center
of mass of the string. For a given energy
$E$, there is a critical impact parameter
$b_c$ which determines wether the string
will fall or not into the black hole. From
our results here for $L=0$ (see
table 1), and for $L\not=0$, we can draw
the following picture: For large impact
parameters
$b>b_c$, the transversal, $\Xi^i$, and temporal, $\Xi^+$, modes
will be stable, while the radial modes, $\Xi^-$, begin to
suffer instabilities. For small impact
parameters, $b<b_c$, the string will fall
into the black hole and, for a Reissner-
Nordstr\o m background, also the transversal
 coordinates suffer instabilities.

It can be noticed that in cosmological inflationary backgrounds,
for which unstable string behavior appears when $(\tau-\tau_0)\to0$, the
string coordinates $\eta^i$, remain bounded. In the black hole
cases, all the characteristic features of string instabilities
appear, but, in addition, the string coordinates $\eta^i$ become
unbounded near the $r=0$ singularity. This happens to be a typical
behavior of strings near spacetime singularities, describing the fact
that the string is trapped by it.

\bigskip\bigskip\bigskip
\noindent
{\it Acknowledgements}

\noindent
The authors thank the Ettore Majorana Center at Erice, where part of
this work was done, for kind hospitality.
C.O.L. thanks the Alexander von Humboldt
 Foundation and the Directorate General
 for Science Research and Development
 of the Commission of the European
 Communities for partial financial support.

\vfill\eject
\references
\doublespace
\bigskip

\refis{vs88} H.J.de Vega and N.S\'anchez,
 Nucl.Phys., B309, 552 and 577 (1988).

\refis{vs87} H.J.de Vega and N.S\'anchez,
Phys.Lett., B197, 320 (1987).

\refis{sv90} N.S\'anchez and G.Veneziano,
 Nucl.Phys., B333, 253 (1990).

\refis{gsv} M.Gasperini, N.S\'anchez and G.Veneziano, Int.J.Mod.Phys.,
A6, 3853 (1991) and Nucl. Phys., B364, 365 (1991).

\refis{gr} I.S.Grandshteyn and I.M.Ryzhik,
"Tables of integrals, Series and Products",
Academic Press, New York (1965).

\refis{rep} W.Israel, in ``Black Hole
Physics", V. De Sabbata and Z. Zhang Eds.,
NATO ASI series, Vol. 364 (1992), p 147.

\refis{v91} N.S\'anchez (unpublished).
G.Veneziano, Phys.Lett., B265, 287 (1991).

\refis{LS93} C.O.Lousto and N.S\'anchez,
in preparation.

\refis{vs91} H.J.de Vega and N.S\'anchez, Phys. Rev., D45, 2783 (1992).

\refis{vms} H.J.de Vega, M.Ram\'on-Medrano and N.S\'anchez, LPTHE - Paris
and DEMIRM - Meudon preprint (1992).

\refis{g92} M.Gasperini, Gen.Rel.Grav., Vol.24, 219 (1992).

\refis{prd92} C.O.Lousto and N.S\'anchez, Phys. Rev., D46, No.10 (1992).

\endreferences

\vfill\eject

\bgroup
\bigskip\bigskip
\bigskip\bigskip

{\bf Table 1} Regimes of string stability in black hole and De Sitter
spacetimes:  Here stable means well behaved string fluctuations
and the usual oscillatory behavior with real frequencies. Unstable
behavior corresponds to unbounded amplitudes $(\Sigma^\pm,\Sigma^i)$
with the emergence of non oscillatory behavior or imaginary
frequencies, accompanied by the infinite string stretching of the
proper string length. $\Sigma^+$,$\Sigma^-$ and $\Sigma^i$, $(i=2,...,
D-1)$, are the temporal, radial and angular (or transverse) string
components respectively.

\bigskip\bigskip
\bigskip\bigskip
\bigskip\bigskip

\def\smallrule{\makeline{\II\noline\I\thinline\thinline\thinline\thinline}}

\table{15cm}{
      \====
      \[  Region        |
          Mode          |
	  Schwarzschild |
	  Reissner-Nordstr\o m |
	  De Sitter    \]
      \====
      \[              |
          $\Xi^{+}$   |
	  stable      |
	  stable      |
	  stable     \]
       \smallrule
       \[ $q^R \to 0$  |
          $\Xi^{-}$    |
	  unstable   |
	  unstable   |
	  unstable  \]
       \smallrule
       \[            |
          $\Xi^{i}$  |
	  stable     |
	  unstable   |
	  unstable  \]
	\====
       \[            |
	  $\Xi^{+}$  |
	  stable     |
	  stable     |
	  stable    \]
       \smallrule
       \[ $q^R \to q_H^R$   |
          $\Xi^{-}$         |
          unstable          |
	  unstable / stable |
          unstable         \]
       \smallrule
       \[            |
          $\Xi^{i}$  |
	  stable     |
	  stable     |
	  unstable  \]
       \====
       \[            |
          $\Xi^{+}$  |
	  stable     |
	  stable     |
	  stable    \]
       \smallrule
       \[ $q^R \to \infty$  |
          $\Xi^{-}$  |
          unstable   |
	  unstable   |
	  unstable  \]
       \smallrule
       \[            |
          $\Xi^{i}$  |
	  stable     |
	  stable     |
	  unstable  \]
       \====  }

\end